\documentclass[universe,article,accept,moreauthors,pdftex]{Definitions/mdpi}
\firstpage{1} 
\pubvolume{1}
\issuenum{1}
\articlenumber{0}
\pubyear{2022}
\copyrightyear{2022}
\externaleditor{Academic Editor: 
} 
\usepackage[normalem]{ulem} 
\usepackage{soul}

\datereceived{13 April 2022} 
\dateaccepted{11 July 2022} 
\datepublished{} 
\hreflink{https://doi.org/} 

\usepackage[version=4]{mhchem}
\usepackage{enumitem}
\usepackage{accents}
\newcommand*{\dt}[1]{%
  \accentset{\mbox{\large\bfseries .}}{#1}}

\Title{Low Density Neutron Star Matter with Quantum Molecular Dynamics: The Role of Isovector Interactions}

\TitleCitation{Low Density Neutron Star Matter with Quantum Molecular Dynamics: The Role of Isovector Interactions}

\Author{Parit Mehta 
 $^{1}$*,  {Rana Nandi} 
 $^{2}$,  {Rosana de Oliveira Gomes} 
 $^{3,}$,  {Veronica Dexheimer} 
 $^{4}$ and  {Jan Steinheimer} 
 $^{5}$}

\AuthorNames{Parit Mehta, Rana Nandi, Rosana de Oliveira Gomes, Veronica Dexheimer and Jan Steinheimer}

\AuthorCitation{Mehta, P.; Nandi, R.; de Oliveira Gomes, R.; Dexheimer, V.; Steinheimer, J. }

\address{%
$^{1}$ \quad I. Physikalisches Institut, Universität zu Köln, Zülpicher Str. 77, 50937 Köln, Germany; mehta@ph1.uni-koeln.de\\
$^{2}$ \quad  Polba Mahavidyalaya, Hooghly, West Bengal 712148, India; nandi@fias.uni-frankfurt.de\\
$^{3}$ \quad Hakom Time Series Gmbh, Lemböckgasse 61, 1230 Vienna, Austria; rosana.gomes@hakom.at  
\\
$^{4}$ \quad Department of Physics, Kent State University, Kent, OH 44242, USA; vdexheim@kent.edu\\
$^{5}$ \quad Frankfurt Institute for Advanced Studies, Ruth-Moufang-Str. 1, 60438 Frankfurt am Main, Germany; steinheimer@fias.uni-frankfurt.de
}
\corres{Correspondence:  {mehta@ph1.uni-koeln.de} 
}

\abstract{The effect of isospin-dependent nuclear forces on the inner crust of neutron stars is modeled within the framework of Quantum Molecular Dynamics (QMD). To successfully control the density dependence of the symmetry energy of neutron-star matter below nuclear saturation density, a mixed vector-isovector potential  is introduced. This approach is inspired by the baryon density and isospin density-dependent repulsive Skyrme force of asymmetric nuclear matter. In isospin-asymmetric nuclear matter, the system shows nucleation, as nucleons are arranged into shapes resembling nuclear pasta. The dependence of clusterization in the system on the isospin properties is also explored by calculating two-point correlation functions.  We show that, as compared to previous results that did not involve such mixed interaction terms, the energy symmetry slope $L$ is successfully controlled by varying the corresponding coupling strength. Nevertheless, the effect of changing the slope of the nuclear symmetry energy $L$ on the crust-core transition density does not seem significant. To the knowledge of the authors, this is the first implementation of such a coupling in a QMD model for isospin asymmetric matter, which is relevant to the inner crust of neutron and proto-neutron stars.}

\begin{document}
\section{Introduction}

Matter in  neutron stars presents the largest densities achieved in the Universe, making their equation of state (EOS) hard to determine. Seeking the EOS of neutron-star matter (NSM) is a flourishing field of interest due to the presence of neutron rich matter with magnetic fields that can be larger than $\mathrm{10^{12}}$ G with the possibility of exotic particles, and~a phase transition to deconfined quark matter. The~crust of a neutron star contains nuclei embedded in a sea of electrons. As~the density increases from the surface of the neutron star towards its core, these nuclei undergo a neutronization process, eventually reaching a state of high neutron to proton asymmetry, which is followed by a transition to uniform nuclear matter at the core. Since matter above nuclear saturation density is unattainable in terrestrial conditions (except in heavy ion collisions with larger temperatures), neutron stars are considered to hold the key to the mysteries of dense nuclear~matter. 

Several approaches have been employed to study the properties of nuclear matter in the context of neutron stars. One of the prominent methods is Quantum Molecular Dynamics (QMD), which allows for the incorporation of competing nuclear forces of attraction and repulsion in dynamical simulations. QMD as a framework for simulating heavy-ion collisions was proposed by J. Aichelin and H. Stöcker~\cite{Aichelin1986QuantumCollisions}. Until~then, nuclear matter simulations were only possible microscopically through one-body models, such as the Vlaslov--Uehling--Uhlenbeck (VUU) theory, and~macroscopically by fluid dynamical models~\cite{Peilert1994PhysicsCollisions,Bohnet1991MultifragmentationThreshold}. QMD combines classical molecular dynamics with quantum corrections, the~most important of which is the Pauli principle. Peilert~et~al.~\cite{Peilert1991ClusteringDensities} used QMD for the first time to simulate clustering in nuclear matter at sub-saturation densities. They performed uniform nuclear matter simulations with nucleons, which were  sampled only in momentum space, for~the density range $0<\rho<2\rho_0$ (where $\rho_0$ is the nuclear saturation density). These were then compared with simulations where nucleons were free to move in position space, showing a decrease in binding energy per nucleon ($E/A$), for~the latter case, of~about 8 MeV towards a more bound system for sub-saturation densities at a near-zero temperature. In~the same work, the~authors also took snapshots of simulated nuclear matter for different mean densities below $\rho_0$, which was useful to visualize clustered matter at $\rho = 0.1\rho_0$, but~did not help deduce the properties of single clusters (unless a computationally expensive time average of many simulations could be done).     

Later, results for sub-saturation density nuclear matter at zero temperature were published by Maruyama~et~al.~\cite{Maruyama1998QuantumDensity}, where the the number of nucleons was significantly expanded (by $\approx$4 times) in the simulated infinite nuclear matter system. In~addition to partially observing transient shapes like holes, slabs, and~cylinders in clustered nuclear matter, they also extended the calculations to asymmetric nuclear matter, and~obtained similar clusterization effects. This is necessary to evaluate the properties of NSM, which is highly asymmetric at saturation and sub-saturation densities. Further improvements to NSM simulations were made by Watanabe~et~al.~\cite{Watanabe2003StructureDynamics} by implementing larger relaxation time scales and analyses of spatial distribution of nucleons. In~a similar analysis, utilizing the Indiana University Molecular Dynamics framework, Sagert~et~al.~\cite{Sagert2016QuantumSimulations} have shown nuclear pasta through similar 3D Skyrme Hartree--Fock (SHF) simulations. Recently, Schramm and Nandi~\cite{Nandi2016LowEnergy} studied the asymmetry dependence of the transition density from asymmetric to homogeneous nuclear matter in the inner crust using~QMD. 

 In this article, the~asymmetry dependence shown by R. Nandi and S. Schramm~\cite{Nandi2016LowEnergy} is modified to have better control on the symmetry energy slope ($L$). The~inspiration is taken from the coupling of omega ($\omega$) and rho ($\rho$) meson fields in the Relativistic Mean-Field (RMF) theory. The~model is first applied to isospin chains of finite nuclei, and~then to nuclear matter at $\rho_0$. Symmetry energy at saturation density is re-evaluated along with its slope $L$. The~primary aim of this work is to successfully control the density dependence of symmetry energy, and~of pure neutron matter, by~calibrating the $\omega-\rho$ type coupling according to established constraints. The~expected clustering of nuclear matter at densities $\approx$$0.1\rho_0$ is also~addressed.           

The structure of the article is as follows: the general formalism is outlined in Section~\ref{sec2}. Then a study of parameters of different strengths of the $\omega-\rho$ coupling in elucidated in Section~\ref{sec3}. The~conclusions  are presented in Section~\ref{sec4}, along with an outlook for the model under study.

\section{Formalism}\label{sec2}
\subsection{The Canonical Formalism: Hamiltonian and Equations of~Motion}

Quantum Molecular Dynamics (QMD) is a model used to accomplish dynamical simulations of nuclear matter by incorporating correlation effects between the constituents of the simulated N-body system. Peilert~et~al.~\cite{Peilert1991ClusteringDensities} studied non-uniformities that give rise to clustering in nuclear matter. A~model based on QMD for heavy-ion collisions through an N-body approach was proposed as early as the late 1980s (Aichelin and Stöcker~\cite{Aichelin1986QuantumCollisions}).     
 The reader can refer to Ref.~\cite{Aichelin1991QuantumCollisions} for a thorough review of the method and its theoretical background. A~brief insight into the working of QMD and the relevance to this project is provided in this section based on a review by Maruyama~et~al.~\cite{Maruyama2012MolecularMatter}.
 
 In a Classical Molecular Dynamics (CMD) simulation of nucleons, particles are simulated as solid elastic spheres, and~their motion is governed by Newton's equations of motion. Inter-particle potentials quantify the force experienced by a particle, given the positions of other particles. QMD introduces quantum behavior to the system of nucleons by including the following modifications: 
 
 \begin{enumerate}
     \item[(a)] In QMD for nuclear matter, a~nucleon is represented by a fixed-width Gaussian wavepacket in the form of a single particle wave function
\begin{equation}\label{eq:nucleon1}
        \psi({\bf{r_i}}) = \frac{1}{(2\pi C_W)^{3/4}} exp\bigg(-\frac{({\bf{r}} - {\bf{R_i}})^2}{4C_W}+i{\bf{r}}\cdot {\bf{P_i}}\bigg) ,
     \end{equation}
 with $\bf{R_i}$ and $\bf{P_i}$ as the centers of position and momentum of the wave packet, respectively. $C_W$ denotes the width of the wave packet. The~motion of the wave packet or 'nucleon' is determined by forces derived from inter-particle potentials in the QMD Hamiltonian. The~total wave function of the N-nucleon system is obtained through a direct product given by
\begin{equation}\label{eq:nucleon2}
       \Psi({\bf{r}}) = \prod_i^N \psi(\bf{r_i}),
    \end{equation}     
     \item[(b)] The nucleon wavefunctions are not anti-symmetrized to explicitly manifest fermionic characteristics. As~a result, the~energy states violate the Pauli principle, as~they all attain minimum energy. This problem was addressed phenomenologically (see the review in Ref.~\cite{Maruyama2012MolecularMatter} for further references) by mimicking the Pauli principle through a repulsive 2-body potential called the Pauli potential ($V_{Pauli}$). The~potential effectively repels nucleons with the same spin and isospin from coming close in phase space, since it is a function of both distance in coordinate and momentum space. In~the ground state, nucleons have non-zero momentum values and do not all exist in the lowest energy state. 
 \end{enumerate}
 
 The Hamiltonian of the nucleon-nucleon interaction is given by Ref.~\cite{Sonoda2008PhaseCores}
\begin{equation}\label{eq:hamiltonian1}
\mathcal{H}=K+V_{\text { Pauli }}+V_{\text { Skyrme }}+V_{\text { sym }}+V_{\mathrm{MD}}+V_{\text { Coul }},
\end{equation}
where $K$ is the kinetic energy, $V_{Pauli}$ is the  {Pauli} 
 potential, $V_{Skyrme}$ is the potential similar to Skyrme like interactions, $V_{sym}$ is the isospin dependent potential, $V_{MD}$ is the momentum dependent potential, and~$V_{Coul}$ is the Coulomb potential. The~expressions for the potential and kinetic terms are
\begin{align}
K &= \sum_{i}\frac{\bf{{P}_{i}^{2}}}{2 m_{i}}~, \label{eq:hamiltoniankin}\\
V_{\text { Pauli }} &=\frac{C_{\mathrm{P}}}{2}\left(\frac{1}{q_{0} p_{0}}\right)^{3} \sum_{i, j( \neq i)} \exp \left[-\frac{\left(\bf{{R}_{i}-{R}_{j}}\right)^{2}}{2 q_{0}^{2}}-\frac{\left(\bf{{P}_{i}-{P}_{j}}\right)^{2}}{2 p_{0}^{2}}\right] \delta_{\tau_{i} \tau_{j}} \delta_{\sigma_{i} \sigma_{j}}\label{eq:paulipotential}~, \\
V_{\text { Skyrme }} &= \frac{\alpha}{2 \rho_{0}} \sum_{i, j( \neq i)} \rho_{i j}+\frac{\beta}{(1+\theta) \rho_{0}^{\theta}} \sum_{i}\left[\sum_{j( \neq i)} \tilde{\rho}_{i j}\right]^{\theta}\label{eq:hamiltonianskyrme}~, \\
V_{\mathrm{Sym}} &=\frac{C_{\mathrm{s}}
}{2 \rho_{0}} \sum_{i, j( \neq i)}\left(1-2\left|\tau_{i}-\tau_{j}\right|\right) \rho_{i j}~,
\label{eq:hamiltoniansym} 
\\ 
V_{\mathrm{MD}} &=\frac{C_{\mathrm{ex}}^{(1)}}{2 \rho_{0}} \sum_{i, j( \neq i)} \frac{1}{1+\left[\frac{ \bf{{P}_{i}-{P}_{j}}}{\mu_{1}}\right]^{2}} ~\rho_{i j}+\frac{C_{\mathrm{ex}}^{(2)}}{2 \rho_{0}} \sum_{i, j( \neq i)} \frac{1}{1+\left[\frac{ \bf{{P}_{i}-{P}_{j}}}{\mu_{2}}\right]^{2}} ~\rho_{i j} \label{eq:hamiltonianmd}~,\\
V_{\mathrm{Coul}} &=C_{coul}\frac{e^{2}}{2} \sum_{i, j( \neq i)}\left(\tau_{i}+\frac{1}{2}\right)\left(\tau_{j}+\frac{1}{2}\right) \iint d^{3} {\bf{r}}\ d^{3} {\bf{r}^{~\prime}} \frac{1}{\left| {\bf{{r}}- {\bf{r}^{~\prime}}}\right|} \rho_{i}( {\bf{r}}) \rho_{j}( {\bf{r}^{~\prime}})~,\label{eq:hamiltoniancoulomb} 
\end{align}\\where the nucleon mass, spin, and~isospin are represented by $m_i$, $\sigma_i$ and $\tau_i$, respectively. The~values of the parameters are listed in Table~\ref{table:parameterset}. 

\begin{table}[H]
\caption{ {Parameter} 
 set for nucleon-nucleon interaction (values from Ref.~\cite{Maruyama1998QuantumDensity} parameterized to reproduce properties of the ground states of the finite nuclei and saturation
properties of the nuclear matter). The~parameters are optimized to give $E/A$ $\approx$ $-$16 MeV for symmetric nuclear matter at saturation $\rho_0$. }
\label{table:parameterset}
%
\newcolumntype{C}{>{\centering\arraybackslash}X}
\begin{tabularx}{\textwidth}{CC}
\noalign{\hrule height 1pt}
\rowcolor{black!30}\textbf{Parameter} & \textbf{Value}\\
\noalign{\hrule height 0.5pt}

$C_P$ (MeV) & 207 \\ 

\rowcolor{black!15}$p_0$ (MeV/c) & 120  \\

$q_0$ (MeV) & 1.644 \\

\rowcolor{black!15}$\alpha$ (MeV) &$-$92.86 \\

$\beta$ (MeV) & 169.28 \\

\rowcolor{black!15}$\theta$ & 1.33333 \\

$C_{ex}^{(1)}$ (MeV) & $-$258.54 \\

\rowcolor{black!15}$C_{ex}^{(2)}$ (MeV) & 375.6 \\

$\mu_1$ (fm$^{-1}$) & 2.35 \\

\rowcolor{black!15}$\mu_2$ (fm$^{-1}$) & 0.4 \\

$\rho_0$ (fm$^{-3}$) & 0.165 \\
\rowcolor{black!15} $C_S$ (MeV) & 25 \\
 $C_W$ (fm$^{2}$) & 2.1 \\
\rowcolor{black!15} $C_{coul}$ & 0 or 1\\
\bottomrule
\end{tabularx}
\end{table}

 The overlap between single nucleon densities $\rho_{i j}$ and $\tilde{\rho}_{i j}$, which depends on positions ${\bf{R_i}} \, \text{and} \, {\bf{R_j}}$, is calculated as
\begin{equation}\label{eq:nucleonoverlap1}
\rho_{i j} \equiv \int d^{3} {\bf{r}} \rho_{i}({\bf{r}}) \rho_{j}({\bf{r}}), \quad \tilde{\rho}_{i j} \equiv \int d^{3} {\bf{r}} \tilde{\rho}_{i}({\bf{r}}) \tilde{\rho}_{j}({\bf{r}})~,
\end{equation}\\where the single nucleon densities are given by
\begin{equation}\label{eq:singlenucleondens}
\begin{aligned} \rho_{i}({\bf{r}}) &=\left|\psi_{i}({\bf{r}})\right|^{2}=\frac{1}{\left(2 \pi C_{W}\right)^{3 / 2}} \exp \left[-\frac{\left({\bf{r}}-{\bf{R}}_{i}\right)^{2}}{2 C_{W}}\right]~, \\ \tilde{\rho}_{i}({\bf{r}}) &=\frac{1}{\left(2 \pi \tilde{C}_{W}\right)^{3 / 2}} \exp \left[-\frac{\left({\bf{r}}-{\bf{R}_{i}}\right)^{2}}{2 \tilde{C}_{W}}\right]~, \\ 
\end{aligned}
\end{equation}\\along with the modified width
\begin{equation}\label{eq:modwidth}
\tilde{C}_{W}=\frac{1}{2}(1+\theta)^{1 / \theta} C_{W}~, 
\end{equation}
which is calculated in this form to incorporate the effect of density-dependent term in Equation~(\ref{eq:hamiltonianskyrme}) (see Section II.B. of Ref.~\cite{Maruyama1998QuantumDensity} for details).  
 
\subsection{Vector-Isovector Interaction~Formalism}
 
 As the numerical model to simulate nuclear matter in conditions pertaining to neutron star crusts has been outlined above, we now move on to the introduction of a nucleon-nucleon interaction potential based on the RMF $\omega-\rho$ vector~interaction. 
 
C. J. Horowitz and J. Piekarewicz~\cite{Horowitz2001Neutron208Pb,Carriere:2002bx} added isoscalar-isovector coupling terms to the non-linear Lagrangian for nuclear matter, and~achieved softening of symmetry energy to control the neutron skin thickness in \ce{^208Pb}. They introduced a RMF Lagrangian density, where the interaction part has the following terms:
\begin{equation}\label{eq:lagrangianomegarho}
\begin{array}{l}
\mathcal{L}_{\text {int }}=\bar{\psi}\left[g_{s} \phi-\left(g_{\mathrm{v}} V_{\mu}+\frac{g_{\rho}}{2} \tau \cdot {\bf{b}}_{\mu}+\frac{e}{2}\left(1+\tau_{3}\right) A_{\mu}\right) \gamma^{\mu}\right] \psi \\
-\frac{\kappa}{3 !}\left(g_{\mathrm{s}} \phi\right)^{3}-\frac{\lambda}{4 !}\left(g_{\mathrm{s}} \phi\right)^{4}+\frac{\zeta}{4 !} g_{\mathrm{v}}^{4}\left(V_{\mu} V^{\mu}\right)^{2}+\frac{\xi}{4 !} g_{\rho}^{4}\left({\bf{b}}_{\mu} \cdot {\bf{b}}^{\mu}\right)^{2} \\
+g_{\rho}^{2} {\bf{b}}_{\mu} \cdot {\bf{b}}^{\mu}\left[\Lambda_{4} g_{\mathrm{s}}^{2} \phi^{2}+\Lambda_{\mathrm{v}} g_{\mathrm{v}}^{2} V_{\mu} V^{\mu}\right]~,
\end{array}
\end{equation}\\where $\psi$ and $\bar{\psi}$ are the baryon and conjugate baryon fields, respectively. V represents the isoscalar $\omega$ meson field, $\phi$ represents the isoscalar-scalar $\sigma$ meson field, isovector $\bf{b}$ is the $\rho$-meson field, and~the photon is denoted by A. $g_{v}$, $g_{s}$, and~$g_{\rho}$ are the respective coupling constants. A~similar Lagrangian with a non-linear $\omega-\rho$ interaction term is employed by F. Grill, H. Pais~et~al.~\cite{Grill2014EquationStars} to study the effect of the symmetry
energy slope parameter, L, on~the profile of the neutron star crust within a Thomas--Fermi~formalism.  

Note that a softening of the symmetry energy around saturation can also be achieved through the use of density dependent couplings (See Figure~4 and the right panel of Figure~2 of Ref.~\cite{Typel:2018cap}).

According to the RMF framework, the~equation of motion for the $\omega $-meson field takes the form,
\begin{equation}\label{eq:omegaexpectation}
m_{\omega}^2\left\langle V_{0}\right\rangle-\sum_{B=n, p} g_{v} \rho_{B}+\frac{\zeta}{3 !} g_{\mathrm{v}}^{4}\left\langle V_{0}\right\rangle^{3}+2g_{\rho}^{2}\Lambda_{\mathrm{v}} g_{\mathrm{v}}^{2} \left\langle b_{0}\right\rangle^2 \left\langle V_{0}\right\rangle=0~,
\end{equation}\\ and similarly for the the $\rho$-meson field:

\begin{adjustwidth}{-\extralength}{0cm}
\begin{equation}\label{eq:rhoexpectation}
m_{\rho}^{2}\left\langle b_{0}\right\rangle-\sum_{B=n, p} g_{\rho} (\rho_{p}-\rho_{n})+\frac{\xi}{3!} g_{\rho}^{4}\left\langle b_{0}\right\rangle^{3}+2g_{\rho}^{2} \Lambda_{4} g_{s}^{2} \left\langle b_{0}\right\rangle\left\langle \phi_{0}\right\rangle^2+ \\ 2g_{\rho}^{2} \Lambda_{\mathrm{v}} g_{\mathrm{v}}^{2} \left\langle b_{0}\right\rangle \left\langle V_{0}\right\rangle^2=0~.
\end{equation}
\end{adjustwidth}

From Equations~(\ref{eq:omegaexpectation}) and (\ref{eq:rhoexpectation}), it is clear that the mean $\omega$ and $\rho$ meson fields depend on the baryon density $\rho_B$ and isospin density $\rho_I=\rho_{p}-\rho_{n}$ , respectively (linearly, if~we ignore higher-order contributions). Equation~(\ref{eq:lagrangianomegarho}) shows how the mixed coupling $\omega-\rho$ potential part of the Lagrangian density depends quadratically on the $\omega$ and $\rho$ meson fields, from~which we can conclude its dependency to be $\sim\rho_B^2\rho_I^2$. Let us approximate it for our QMD model, motivated by the density-dependent repulsive Skyrme potential as in  Equation~(\ref{eq:hamiltonianskyrme}) with a term quadratic in both the $\rho_{B}$, and~in $\rho_{I}$ as
\begin{equation}\label{eq:hamiltonianomegarho}
V_{\omega\rho} = \frac{C_{\omega\rho}}{5\rho_0^4}\sum_{i,k}<\rho_i>^2\,<\tilde{\rho_k}>^2,
\end{equation}
where $<\rho_i>$ and $<\tilde{\rho_k}>$ are the averaged $\rho_{B}$ and $\rho_{I}$ respectively, with~the \linebreak{following expressions:}
\begin{eqnarray}\label{eq:nucleonoverlap2}
<\rho_i> &=&\sum_{j( \neq i)}\rho_{ij}=\sum_{j( \neq i)}\frac{e^{-({\bf R_i}-{\bf R_j})^2/4C_W}}{(4\pi C_W)^{3/2}}\\
<\tilde{\rho_k}>&=&\sum_{l( \neq k)}c_{kl}\rho_{kl}=\sum_{l( \neq k)}(1-2|\tau_k-\tau_l|)\frac{e^{-({\bf R_k}-{\bf R_l})^2/4C_W}}{(4\pi C_W)^{3/2}}.
\end{eqnarray}

The summation needs to be calculated before squaring in Equation~(\ref{eq:hamiltonianomegarho}). A~similar calculation has already been made for the repulsive part of the Skyrme~potential.

The components of force for the $\omega-\rho$ term can be derived from the potential
\begin{equation}
\begin{aligned}
\begin{array}{lllll}
\label{eq:omegarhoforce}
-f_m^x&=&\frac{\partial V_{\omega\rho}}{\partial X_m}\\
     &=&\frac{2C_{\omega\rho}}{5\rho_0^4}\sum_{j,k}\Bigg[\left(<\rho_m> + <\rho_j>\right)\frac{X_m-X_j}{2L}\rho_{mj} <\tilde{\rho_k}>^2\\
     &&+\left(<\tilde{\rho_m}>+<\tilde{\rho_j}>\right)<\rho_k>^2\frac{X_m-X_j}{2L}c_{mj}\rho_{mj}\Bigg]\\
     &=&\frac{2C_{\omega\rho}}{5\rho_0^4}\Bigg[\sum_{k}<\tilde{\rho_k}>^2\left\{\sum_j\rho_{mj}\frac{dX_{mj}}{2L}(<\rho_m>+<\rho_j>) \right\}\\
     &&+\sum_{k}<\rho_k>^2\left\{\sum_j C_{mj}\rho_{mj}\frac{dX_{mj}}{2L}(<\tilde{\rho}_m>+<\tilde{\rho}_j>)\right\}\Bigg]
\end{array}
\end{aligned}
\end{equation}
 where $X_m$ and $X_j$ are the x-coordinates of the centers of the positions of $m-$th and $j-$the particles, respectively. 
\subsection{Modeling of Infinite Systems: Achieving the Ground State~Configuration}
 Different methods can be employed to achieve the ground state configuration of nuclear matter for a given density or temperature. Peilert~et~al.~\cite{Peilert1991ClusteringDensities} calculated $E/A$ values for finite nuclei, and~subsequently studied infinite nuclear matter using a version of the QMD model. They found that nuclear matter simulated at temperatures near $T=0$ MeV showed clustering among nucleons at sub-saturation densities. Later, Maruyama~et~al.~\cite{Maruyama1998QuantumDensity} employed QMD to study the dynamical evolution of nuclear matter into pasta phases. In~this work, we follow the method employed by Maruyama~et~al., obtaining the energy-minimum configuration of nuclear matter by distributing nucleons randomly in phase space, and~then cooling down the system to achieve the minimum energy state of the system. This allows for arbitrary nuclear shapes and incorporates thermal fluctuations, giving an insight into the formation process of such~structures. 
 
  To achieve equilibrium in the nuclear matter system, we use the following equations of motion along with damping factors $\xi_{R}$ and $\xi_{P}$:
\begin{equation}\label{eq:qmdmotion}
\begin{aligned} {\dt{\bf{R}_{i}}} &=\frac{\partial \mathcal{H}}{\partial {\bf{P}_{{i}}}}-\xi_{R} \frac{\partial \mathcal{H}}{\partial \bf{R}_{{i}}}~, \\ \dt{{\bf{P}_{i}}} &=-\frac{\partial \mathcal{H}}{\partial \bf{R}_{{i}}}-\xi_{P} \frac{\partial \mathcal{H}}{\partial \bf{P}_{{i}}}~, \end{aligned}
\end{equation}
 where $\mathcal{H}$ is given by Equation~(\ref{eq:hamiltonian1}) and the factors $\xi_{R}$ and $\xi_{P}$ are adjusted according to the relaxation time scale, with~a fixed value of either 0 or~$-$0.1.
 
 The system is cooled from an initial temperature maintained by the Nosé--Hoover thermostat. The~thermostat introduces additional coordinates and velocities in the Hamiltonian of the system in order to mimic a thermal bath in contact with the system. The~extended Hamiltonian $\mathcal{H}_{Nose}$ appears as
\begin{equation}\label{eq:nosehoover1}
\begin{aligned} \mathcal{H}_{\mathrm{Nose}} &=\sum_{i=1}^{N} \frac{\bf{P}_{i}^{2}}{2 m_{i}}+\mathcal{U}\left(\left\{\bf{R}_{i}\right\},\left\{\bf{P}_{i j}\right\}\right)+\frac{s p_{s}^{2}}{2 Q}+g \frac{\ln s}{\beta} \\ &=\mathcal{H}+\frac{s p_{s}^{2}}{2 Q}+g \frac{\ln s}{\beta}~,  \end{aligned}
\end{equation}
 where $s$ is the additional dynamical variable for time scaling, $p_{s}$ is the momentum conjugate to $s$,~ $\mathcal{U}\left(\left\{\mathbf{R}_{i}\right\},\left\{\mathbf{P}_{i}\right\}\right)=\mathcal{H}-K$ is the potential which depends on both positions and momenta, $Q$ is the thermal inertial parameter corresponding to a coupling constant between the system and thermostat, $g$ is a parameter to be determined as $3 N$ by a condition for generating the canonical ensemble in the classical molecular dynamics simulations, and~$\beta$ is defined as $\beta \equiv 1 / (k_{\mathrm{B}} T_{\mathrm{set}})$ \cite{garcia2006nose,Watanabe2004PhasesDensities}.
 The energy of the nuclear matter system is not conserved, but~$\mathcal{H}_{Nose}$ is . The~most important variables here are $\beta = 1/k_BT_{set}$~, where $T_{set}$ is the desired input temperature, and~$Q \approx 10^8$ MeV $\mathrm{(fm/c)^2}$. More  details can be found in Refs.~\cite{Watanabe2004PhasesDensities,Nandi:2017aqq} and sources therein.
%
 
 At sub-saturation densities, local minima may take place around the actual global energy-minimum value of the ground state that the damping coefficients lead to if not chosen carefully. The~simulation results should be checked to avoid local energy minima by repeating the cooling procedure.  
{%

%


\section{Results}\label{sec3}
\subsection{Simulation~Procedure}
Using the theoretical framework established in the previous chapters, the~QMD simulation of a system of neutrons and protons is carried out. The~final temperature after cooling down from a finite temperature was set to 0, so as to imitate the conditions in a neutron star's inner~crust. 

A cubic box confines the nucleons. The~size of the box is determined by the number of nucleons $N$ and average density $\rho_{av}$. Periodic boundary conditions are imposed and the motion of nucleons is imitated across 26 cells surrounding the central primitive cell. The~value of $N$ is set to 1024, such that for homogeneous symmetric nuclear matter the number for protons/neutrons with spin up and protons/neutrons with spin down is equal (proton fraction $Y_p$ = 0.5 with 512 particles each of protons and neutrons). Hence, there is no magnetic polarization. Electrons are treated as a uniform background gas that makes the system charge~neutral.

The nucleons are initially distributed randomly in phase space. The~system is brought to thermal equilibrium at $T$ = 20 MeV for about 1000 fm/c. The~system, initially kept at a constant temperature by the Nosé--Hoover thermostat, is slowly cooled down in accordance with the equations of motion (Equation (\ref{eq:qmdmotion})), until~the temperature is 0. To~attain the ground state configuration, the~simulation requires about 1--2 days of computation time to reach $10^4$ fm/c when carried out on the Goethe-HLR CPU cluster at Goethe-University Frankfurt. The~computer code for the simulations in this project was first used for QMD calculations in Ref.~\cite{Nandi2016LowEnergy}.        

The set of values for the parameters used in interaction potentials constituting the Hamiltonian Equation~(\ref{eq:hamiltonian1}) is given in Table~\ref{table:parameterset}. Additionally, the~set of values for the coefficient $C_{\omega\rho}$ of $V_{\omega\rho}$ in Equation~(\ref{eq:hamiltonianomegarho}) are listed in Table~\ref{table:parameterset2}.

\begin{table}[H]
\caption{ {Optimized} 
 values for coefficient of $V_{\omega\rho}$.}
\label{table:parameterset2}
%
\newcolumntype{C}{>{\centering\arraybackslash}X}
\begin{tabularx}{\textwidth}{CC}
\noalign{\hrule height 1pt}

\rowcolor{black!30}\textbf{Set} & \boldmath{$C_{\omega\rho}$} \textbf{(MeV)}\\
\noalign{\hrule height 0.5pt}

 I & 0.02 \\

\rowcolor{black!15}II & 0.01 \\

III & 0.005\\

\rowcolor{black!15}IV & $-$0.01 \\

V & $-$0.02 \\ 
\bottomrule
\end{tabularx}
\end{table}

\subsection{Finite~Nuclei}
We first calculate the binding energies of ground states of a number of finite nuclei and their isotopes. Five different values of the coefficient $C_{\omega\rho}$ are tested. All five reproduce the trend of binding energies per nucleon of various nuclear isotopes, as~can be seen in Figure~\ref{fig:finitebinding}. Individual simulated energy values ($E_{calc}$) deviate from the experimental ($E_{exp}$) counterparts~\cite{nucl-energies-ame2016} by less than 10\% in all cases. Considering a reasonable expectation of accuracy within the QMD model employed in this paper, there is a minor spread in the calculated values. It is clear that varying $C_{\omega\rho}$ does not have a significant impact on the binding energies per nucleon of finite nuclei, which can be explained by the non-dependence of symmetry energy in a finite nucleus to its slope $L$, and~the fact that it rather depends on other parameters: the symmetry energy coefficient at saturation density, ratio of the surface symmetry coefficient to the volume symmetry coefficient, surface stiffness and obviously the mass number of the nucleus (see Refs.~\cite{centelles-nuclear-sym,brown-sym-constraints} and sources therein.)   
The nuclei chosen are heavy with $Z$ larger than 40. Good results for lighter nuclei are not expected, based on the results in Figure~4 of Ref.~\cite{Maruyama1998QuantumDensity}. For~each isotope family, three nuclei are selected with $Y_P$ ranging from 0.3 to 0.5 to analyze the effect of isospin dependent~interactions. 

    \begin{figure}[H]
        \includegraphics[width=12.5 cm]{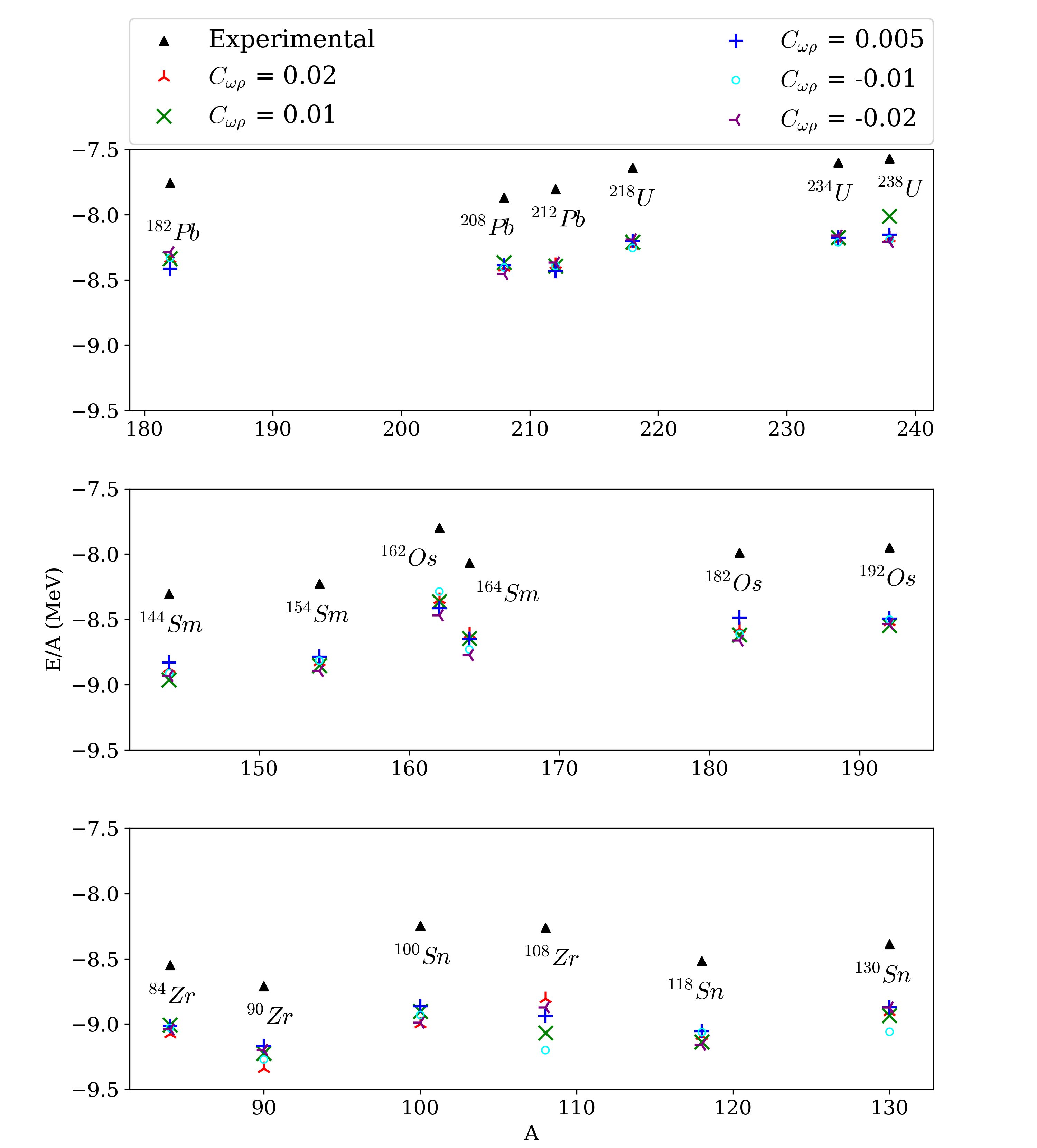}
        \caption{ {Binding} 
 energies per nucleon for three nuclear isotopes each of Zr, Sn, Sm, Os, Pb, and~U obtained from simulation for five different parameter sets listed above the image. The experimental values are taken from AME2016~\cite{nucl-energies-ame2016}.}
        \label{fig:finitebinding}
    \end{figure}

There is an anomaly in the form of binding energies per nucleon being about \linebreak 0.65 MeV too deep compared with experimental values for all nuclei.
Given the realistically achievable accuracy
within a molecular dynamics approach, this deviation is acceptable. Nevertheless, the~model reproduces the overall trends of the binding energies of various nuclei reasonably well, for~all values of $C_{\omega\rho}$.


\subsection{Pure Neutron~Matter}

An important final test of the model is the examination of the behavior of a pure neutron gas at nuclear and sub-nuclear densities. The~energy per nucleon $(E/N)_{n}$ of pure neutron matter affects the densities at which NSM becomes~uniform.
    
    For this case, the~same system is adapted to simulate nuclear matter with $Y_P = 0.0$, i.e.,~1024 neutrons in the primitive cell without protons. The~results for pure neutron matter simulations for nuclear and sub-nuclear densities are shown in Figure~\ref{fig:pureneutron1}. The~density dependence of neutron matter (or the neutron matter EoS) is crucial, as~$E/N$ is an input in the calculation of the symmetry~energy.
\vspace{-6pt}
\begin{figure}[H]
  \includegraphics[width=12.5 cm]{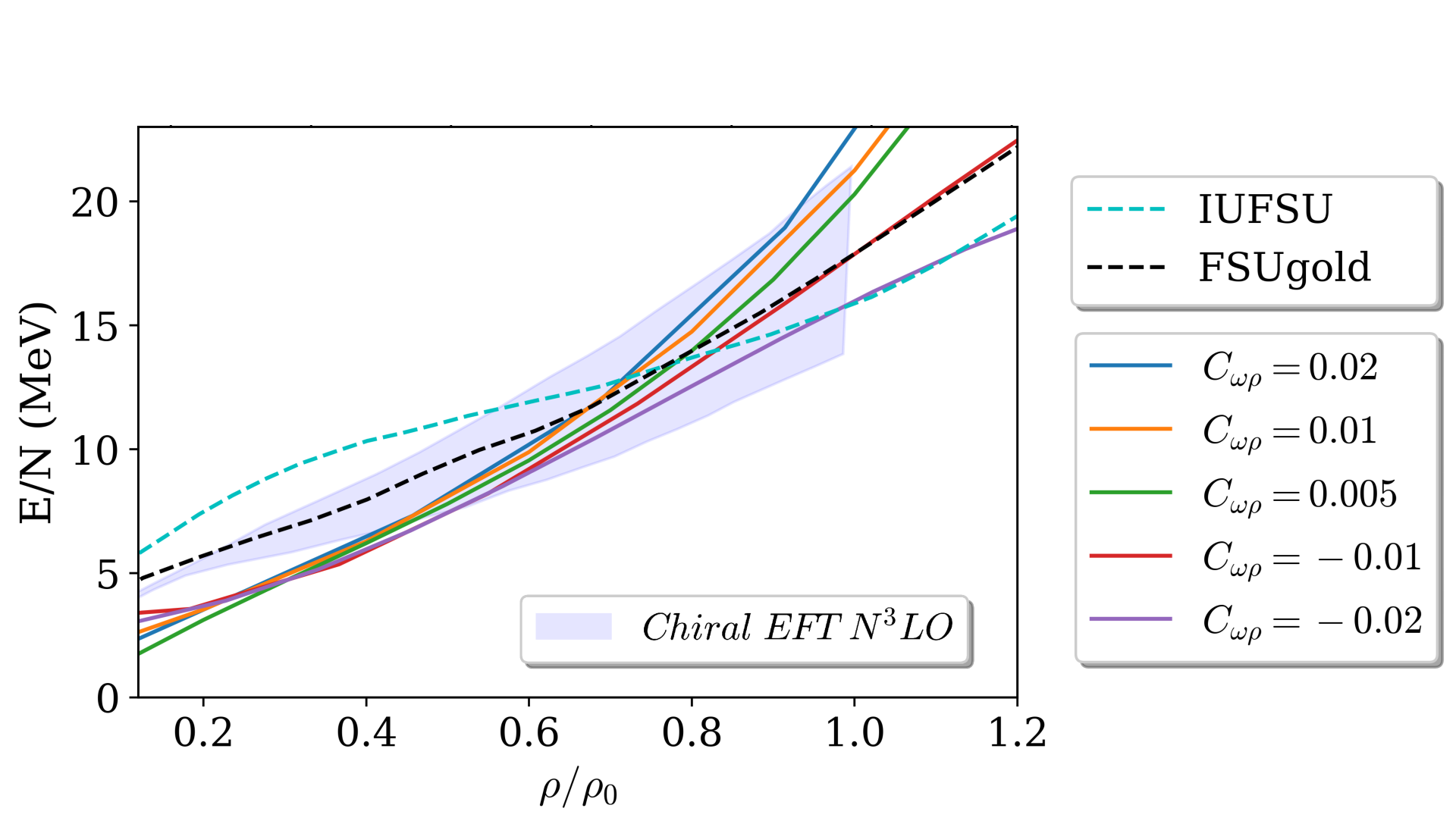} \\
 \caption{ {Energy} 
 per nucleon of pure neutron matter as a function of density for 5 different parameter sets. The~shaded area corresponds to Chiral EFT constraints as provided in Ref.~\cite{Kruger2013NeutronInteractions}. The~RMF models FSUgold and IUFSU, which also include the $\omega-\rho$ interaction, are shown for comparison. Note that here the saturation density $\rho_0=0.165$ $fm^{-3}$.}  \label{fig:pureneutron1}
\end{figure}

     In Figure~\ref{fig:pureneutron1}, the~neutron matter EoSs for different $C_{\omega\rho}$ from the QMD model can be compared with two other non-linear RMF models (IUFSU~\cite{Fattoyev2010RelativisticStars} and FSUgold~\cite{Todd-Rutel2005Neutron-richMatter}), which also include the $\omega-\rho$ coupling. The~shaded area shows the results from Chiral EFT~\cite{Kruger2013NeutronInteractions}, providing robust theoretical constraints for neutron-matter equations of state. For~\linebreak $C_{\omega\rho}$ = 0.02, the~EoS indicates a bit too much repulsion around the nuclear saturation density. For~all values of $C_{\omega\rho}$ at low densities, binding is weaker than expected. In~spite of these issues, all parameter sets with different strengths of the coefficient $C_{\omega\rho}$ appear to be in good qualitative agreement with the~constraints.    
     
    
    The slope of the symmetry energy $L$ and the pure neutron matter EoS are related, shown by Equation~(19) in Ref.~\cite{Sonoda2008PhaseCores}:
\begin{equation}\label{eq:slopeneutron}
        L = 3\rho_0\frac{\partial}{\partial\rho_n}\bigg(\frac{\varepsilon_n}{\rho_n}\bigg)_{\rho_0}~,
    \end{equation}
    where the energy density of pure neutron matter is given by $\varepsilon_n$. The~slope of the neutron-matter EoS decreases as $C_{\omega\rho}$ is lowered, which is consistent with the trend of the $L$ values in Table~\ref{table:symmcalc1}. Therefore, varying the slope (and by extension the strength of the $\omega-\rho$ interaction) has a direct impact on the densities at which neutrons drip out of nuclei, and~consequently on the nuclear pasta phases in~NSM.   
    
 \begin{table}[H]
\caption{ {Symmetry} 
 energies and corresponding slope values (parabolic approximation).}
\label{table:symmcalc1}
%
\newcolumntype{C}{>{\centering\arraybackslash}X}
\begin{tabularx}{\textwidth}{CCCC}
\noalign{\hrule height 1pt}

\rowcolor{black!30}\textbf{Set} & \boldmath{$C_{\omega\rho}$} \textbf{(MeV)} & \boldmath{$S(\rho$)} & \boldmath{$L$}\\
\noalign{\hrule height 0.5pt}

 I & 0.02 & 37.40 & 135.26\\

\rowcolor{black!15} II & 0.01 & 35.63 & 102.71 \\

III & 0.005 & 34.72 & 100.41\\

\rowcolor{black!15}IV & $-$0.01 & 32.23 & 66.38\\

III & $-$0.02 & 30.52 & 48.32\\
\bottomrule
\end{tabularx}
\end{table}   
    
\subsection{Determination  of  Symmetry  Energy  and  Slope~Parameter}\label{sec:3.4}

In a free fermion gas of nucleons, the~expression for energy per particle is
\begin{equation}
    \frac{E}{A}\approx \frac{E}{A} (\beta_{asy} = 0)+E_{sym}\beta_{asy}^2 + ...    ~,
\end{equation}
where $\beta_{asy}$ defined as
\begin{equation}\label{eq:asympar1}
    \beta_{asy} = \frac{n_n - n_p}{n_n + n_p} = \frac{N-Z}{A}~,
\end{equation}
or in terms of proton and neutron densities $\rho_p$ and $\rho_n$,
\begin{equation}\label{eq:asympar2}
    \beta_{asy} = \frac{\rho_n - \rho_p}{\rho}~.
\end{equation}

For an initial determination of $E_{sym}$ and $L$ at saturation density, a~parabolic approximation is applied, such that only the lowest-order non-vanishing term in $\beta_{asy}$ is retained. Rewriting the equation with the approximation gives 
\begin{equation}\label{eq:symmparabolic}
    \frac{E}{A} = \frac{E}{A} (\beta_{asy} = 0)+S(\rho)\beta_{asy}^2~,
\end{equation} 
where $\frac{E}{A} (\beta_{asy} = 0)$ = $(E/A)_0$ is the energy per nucleon of symmetric matter, and~$S(\rho)$ is the nuclear symmetry energy. Keeping the Coulomb interaction switched off, the~simulation is run for many values of $Y_P$ at $\rho_0$ for all $C_{\omega\rho}$ in Table~\ref{table:parameterset2}. The~values for $E/A$ are fitted in Equation \eqref{eq:symmparabolic}, and~$S(\rho_0)$ is obtained as a fit parameter from the plot of energies per nucleon shown in Figure~\ref{fig:symmfit1}.  

The slope parameter $L$ quantifies the density dependence of the symmetry energy, which can be used to practically calculate the possible $L$ values as~\cite{Baldo:2016jhp}
\begin{equation}\label{eq:slopecalc}
    L = 3\rho_0\frac{S(1.1\rho_0) - S(0.9\rho_0)}{1.1\rho_0-0.9\rho_0}~.
\end{equation}

Here, $S(1.1\rho_0)$ and $S(0.9\rho_0)$ are determined with the same procedure as for $S(\rho_0)$ described above. The~obtained values for $S(\rho)$ and $L$ are listed in Table~\ref{table:symmcalc1}.
\vspace{-6pt}
\begin{figure}[H]
    \includegraphics[width=11.5 cm]{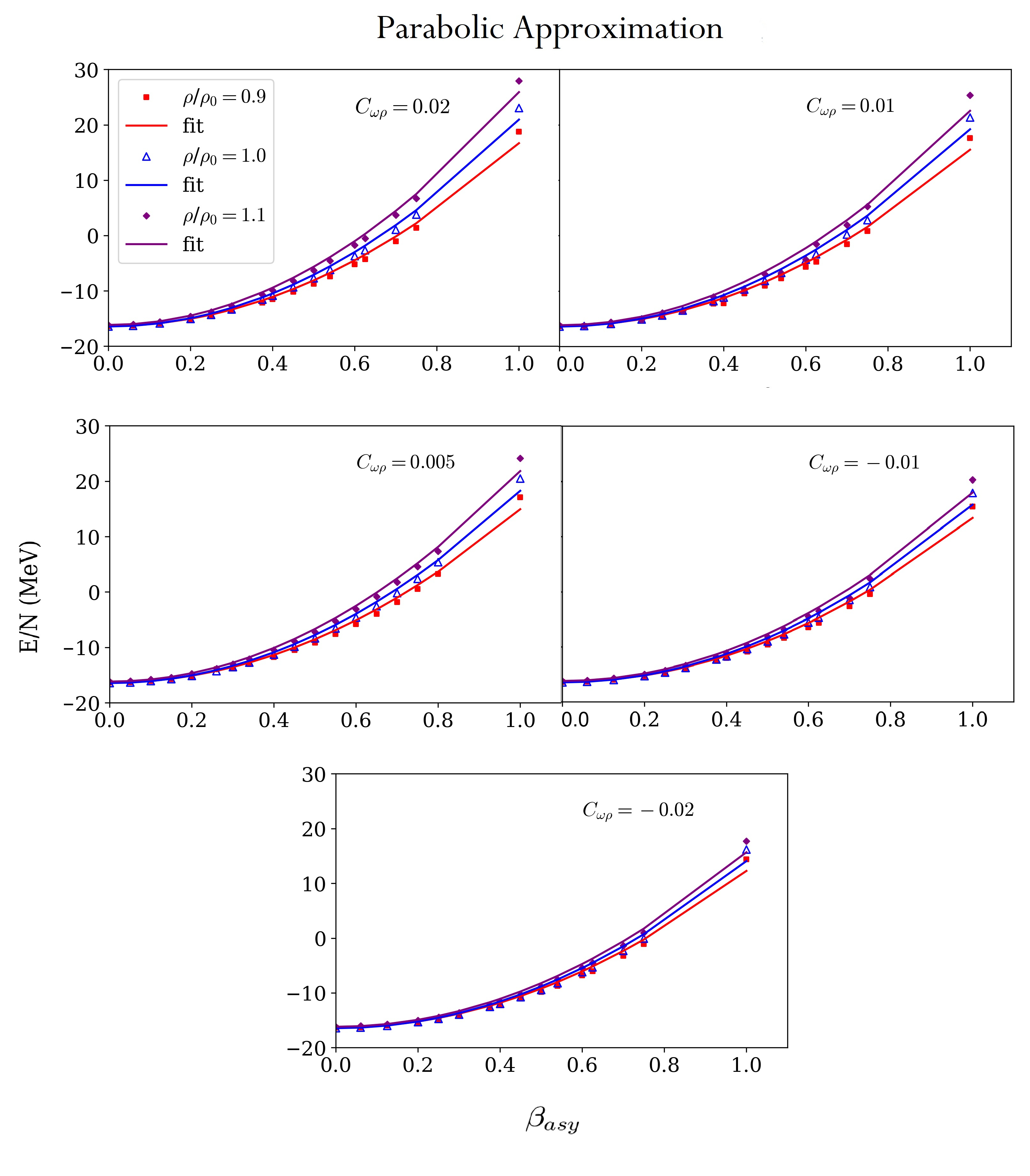}
    \caption{ {Fit} 
 of energy per nucleon vs. neutron excess using Equation \eqref{eq:symmparabolic} for different parameter sets (parabolic approximation). $\beta_{asym}$ is the neutron excess with 1.0 being pure neutron matter, and~0.0 being symmetric nuclear matter. A~list of the corresponding slope values is given in Table~\ref{table:symmcalc1}.}
    \label{fig:symmfit1}
\end{figure}

The calculations discussed above can be improved. Chen~et~al.~\cite{Chen2009Higher-orderMatter}  suggested that the description of the  nuclear matter EoS can be made better by improving on the parabolic approximation. Through a systematic study of isospin dependence of saturation properties of asymmetric nuclear matter, it was concluded that the parabolic approximation produces good results for $\beta_{asy}^2 \leq$ 0.1, but~for higher asymmetries the quartic term should also be included. In~this work, where higher isospin asymmetries are simulated, the~fit using the function in Equation~(\ref{eq:symmparabolic}), as~can be seen in Figure~\ref{fig:symmfit1}, is not satisfactory. The~slope values for $\beta_{asy}^2 >$ 0.1 can therefore be modified by adding a quartic term to Equation~(\ref{eq:symmparabolic}), which now expands to
\begin{equation}\label{eq:symmquartic}
    \frac{E}{A} = \frac{E}{A} (\beta_{asy} = 0)+S_{(2)}(\rho)\beta_{asy}^2+S_{(4)}(\rho)\beta_{asy}^4~,
\end{equation}\\where $S_{(2)}(\rho)$=$S(\rho)$ and $S_{(4)}(\rho)$ is the fourth order term of nuclear symmetry energy. The~binding energies for different $\beta_{asy}$ values are fitted to the Equation~(\ref{eq:symmquartic}) and $S_{(2)}(\rho)$=$S(\rho)$ and $S_{(4)}(\rho)$ are obtained as fitting parameters. A~better fit for energy per nucleon is achieved, as~shown in Figure~\ref{fig:symmfit2}. The~updated values for symmetry energy and slope are listed in Table~\ref{table:symmcalc2}. Figure~\ref{fig:symmenergydens1} shows the density dependence of symmetry energy for 5 different parameter sets. The~difference between results obtained using different parameter sets increases with density due to the quadratic dependence of $V_{\omega-\rho}$ on baryon and isospin densities, being very small for densities below $0.5\rho_0$.
\vspace{-6pt}
\begin{figure}[H]
    \includegraphics[width=11.5 cm]{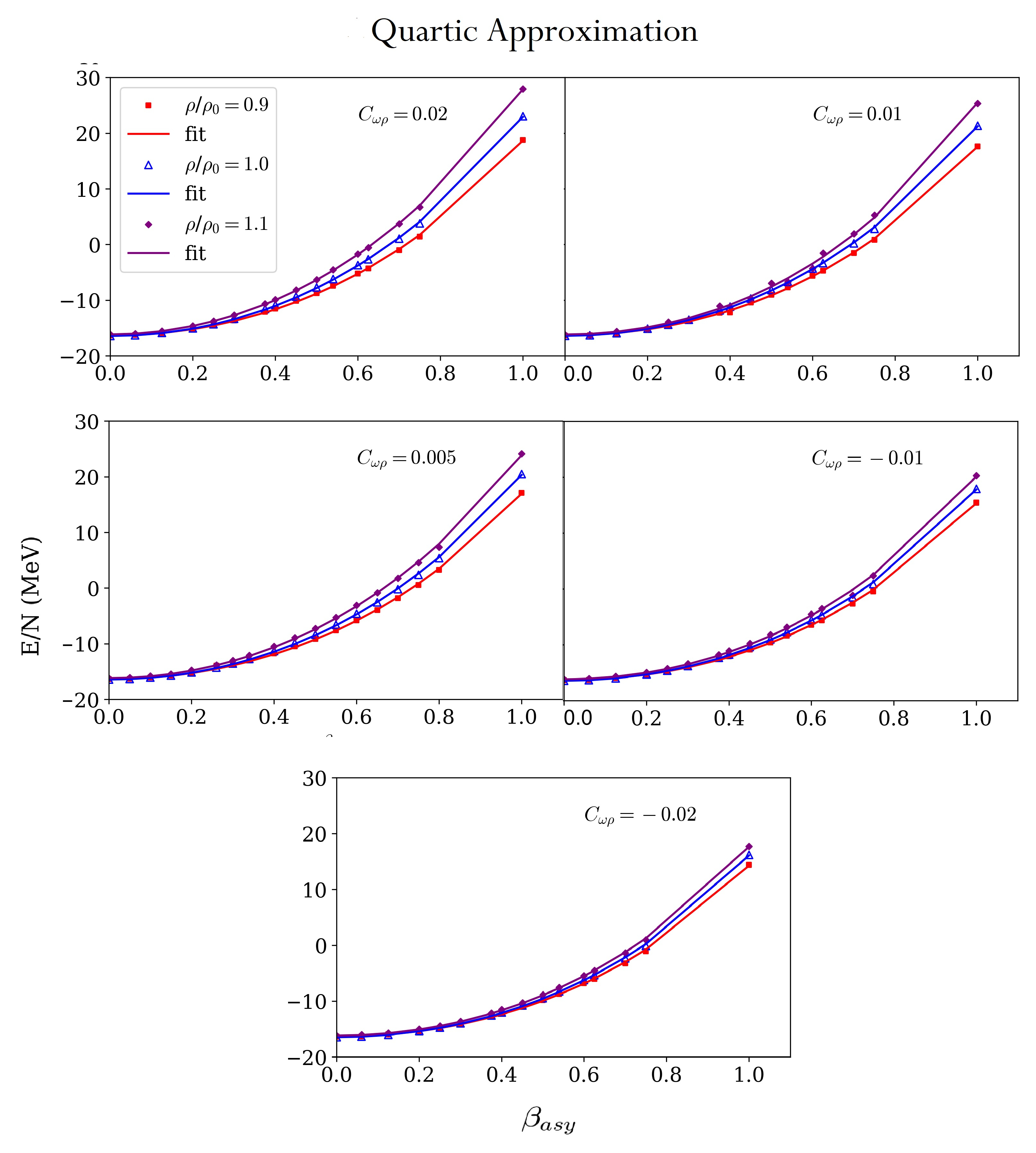}
    \caption{ {Fit} 
 of energy per nucleon vs. neutron excess using Equation \eqref{eq:symmquartic} (quartic approximation) for different parameter sets. $\beta_{asym}$ is the neutron excess with 1 being pure neutron matter and 0 being symmetric nuclear matter. A~list of the corresponding slope values is given in Table~\ref{table:symmcalc2}. This approximation results in better fitting compared to the parabolic approximation in Figure~\ref{fig:symmfit1}.}
    \label{fig:symmfit2}
\end{figure}

\begin{figure}[H]
 \includegraphics [width=8.5 cm]{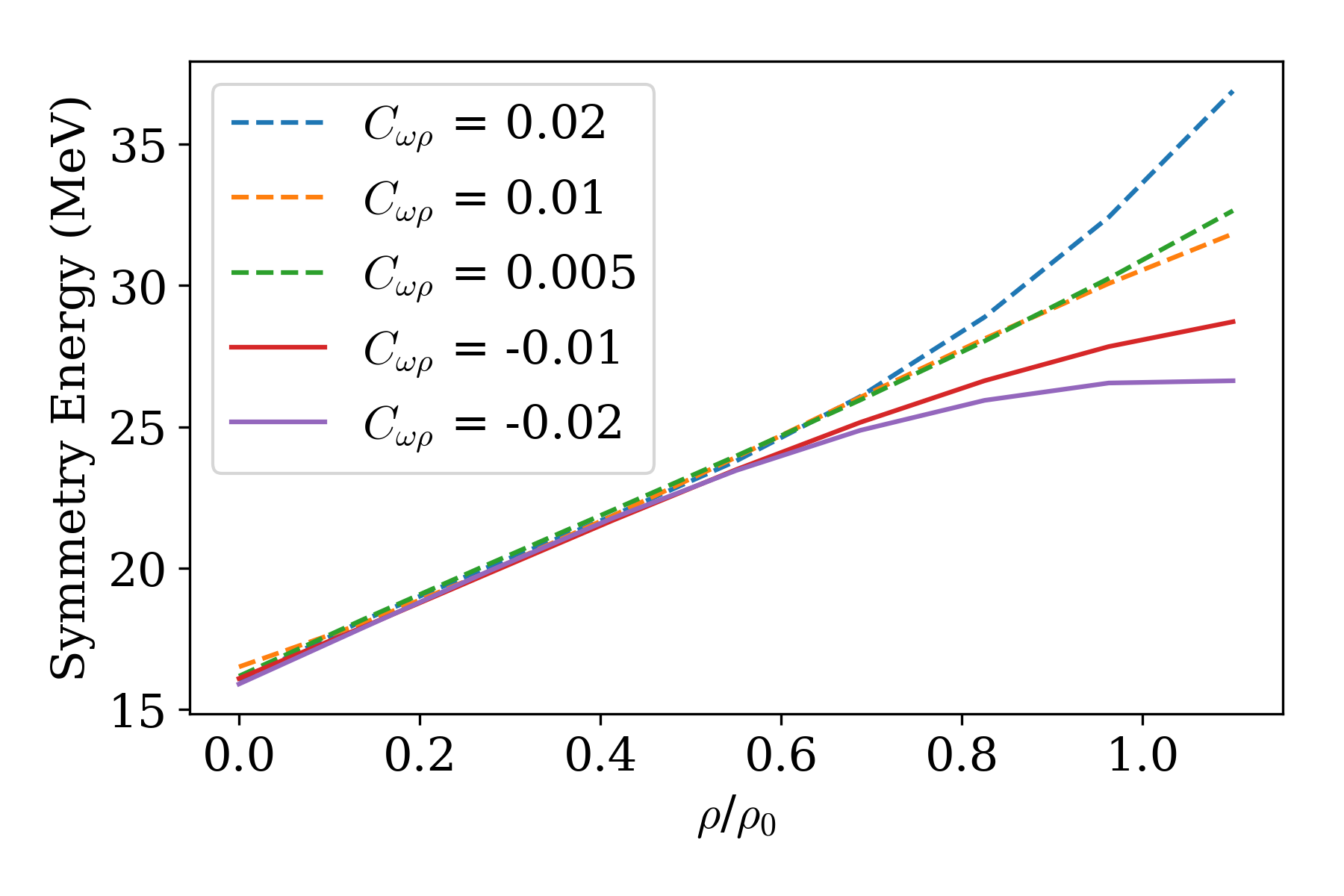}
  \caption{Density dependence of symmetry energy for 5 different parameter sets is shown. The~$\omega-\rho$ term determines the coupling of two vector fields, which are sub-leading at low densities, where the attraction represented by the scalar fields dominates. As~the density increases beyond $\rho_B=0.5$ fm$^{-3}$, the~$\omega-\rho$ effects can clearly be seen in this~figure.}  \label{fig:symmenergydens1}
\end{figure} 

Comparing Tables~\ref{table:symmcalc1} and \ref{table:symmcalc2}, the~effect of adding a fourth order term is a decrease in the symmetry energy in all cases. However, the~decrease in $L$ is not straightforward and is only seen for Sets II, III, and~IV in the quartic case (as compared to the parabolically approximated case). It is clear in Table~\ref{table:symmcalc2} that, with~a decrease in $C_{\omega\rho}$, $L$ can be lowered to optimal values for Sets II, III, VI, and V. However, the~$L$ for Set III was expected to be lower than that for Set II, in~agreement with the trend of decreasing values going from Sets I to V. It appears that this value is indeed much closer to previous results in Ref.~\cite{Nandi2016LowEnergy}, where a symmetry energy of $\approx$29 MeV is associated with an $L$ $\approx$ 92 MeV. This can be interpreted in terms of the strength of the $\omega-\rho$ interaction energy being too low for Set III, which causes the prediction to agree closely with previous results that excluded it. Note that the values of symmetry energy and slope for set I are high when compared with experimental values~\cite{Li:2019xxz,PhysRevLett.126.172503,PhysRevLett.127.232501}.

\begin{table}[H]
\caption{ {Symmetry} 
 energies and corresponding slope values (quartic approximation).}
\label{table:symmcalc2}
%
\newcolumntype{C}{>{\centering\arraybackslash}X}
\begin{tabularx}{\textwidth}{CCCCCC}
\noalign{\hrule height 1pt}

\rowcolor{black!30}\textbf{Set} & \boldmath{$C_{\omega\rho}$} \textbf{(MeV)} & \boldmath{$S_{(2)}(\rho)$} & \boldmath{$S_{(4)}(\rho)$} &  \boldmath{$L_{(2)}$} & \boldmath{$L_{(4)}$}\\ 
\noalign{\hrule height 0.5pt}
 I & 0.02 & 32.81 & 6.56 & 135.52 & $-$0.38\\

\rowcolor{black!15} II & 0.01 & 30.93 & 6.72 & 71.88 & 44.11\\

III & 0.005 & 30.18 & 6.59 & 99.52 & 1.30\\

\rowcolor{black!15}IV & $-$0.01 & 27.56 &  6.68 & 61.32 & 7.24\\

V & $-$0.02 & 25.88 & 6.63 & 49.32 & $-$1.43\\
\bottomrule
\end{tabularx}
\end{table}

\subsection{Nucleon~Distributions}\label{sec:pasta}
Nuclear clustering cannot only occur in NSM, but~also for more isospin symmetric matter as it undergoes the liquid-gas transition. Such matter can be studied, for~example, in~high energy nuclear collisions. In~order to bridge the gap between such studies and NSM, the~proper isospin dependence of the existence and occurrence of the liquid-gas phase separation needs to be understood. In~the following we will show how our model can be used to study the occurrence of clustering of nuclear matter for nuclear matter with proton fractions between $0.3<Y_e<0.5.$ 

The nucleon distribution of nuclear matter at $T=0$ can be visualized in the simulation box. At~every grid point, the~density contribution of each nucleon is added to produce a density map. Since the Coulomb interaction is included, clusterization of nucleons in a lattice-like structure is clearly seen in the~system. 

In Figures~\ref{fig:cluster1}--\ref{fig:cluster3}, clusterization is seen in the system at $T=0$.  At~ every  grid  point  in  the  simulation  box,  the~ density  contribution  of each nucleon gaussian wave packet is added to calculate the density map. In~Figure~\ref{fig:cluster2}, an~increased $C_{\omega\rho}$ decreases the density of clusters, so they seem to break up into smaller clusters of lower densities. This implies that neutrons will drip out at lower densities for $C_{\omega\rho}$ = 0.02 than for 0.01. As~the density is increased 3-folds (shown in Figure~\ref{fig:cluster3}), the~density map morphs into a more interesting~structure. 
\vspace{-6pt}
    \begin{figure}[H]
    \includegraphics[width=7 cm]{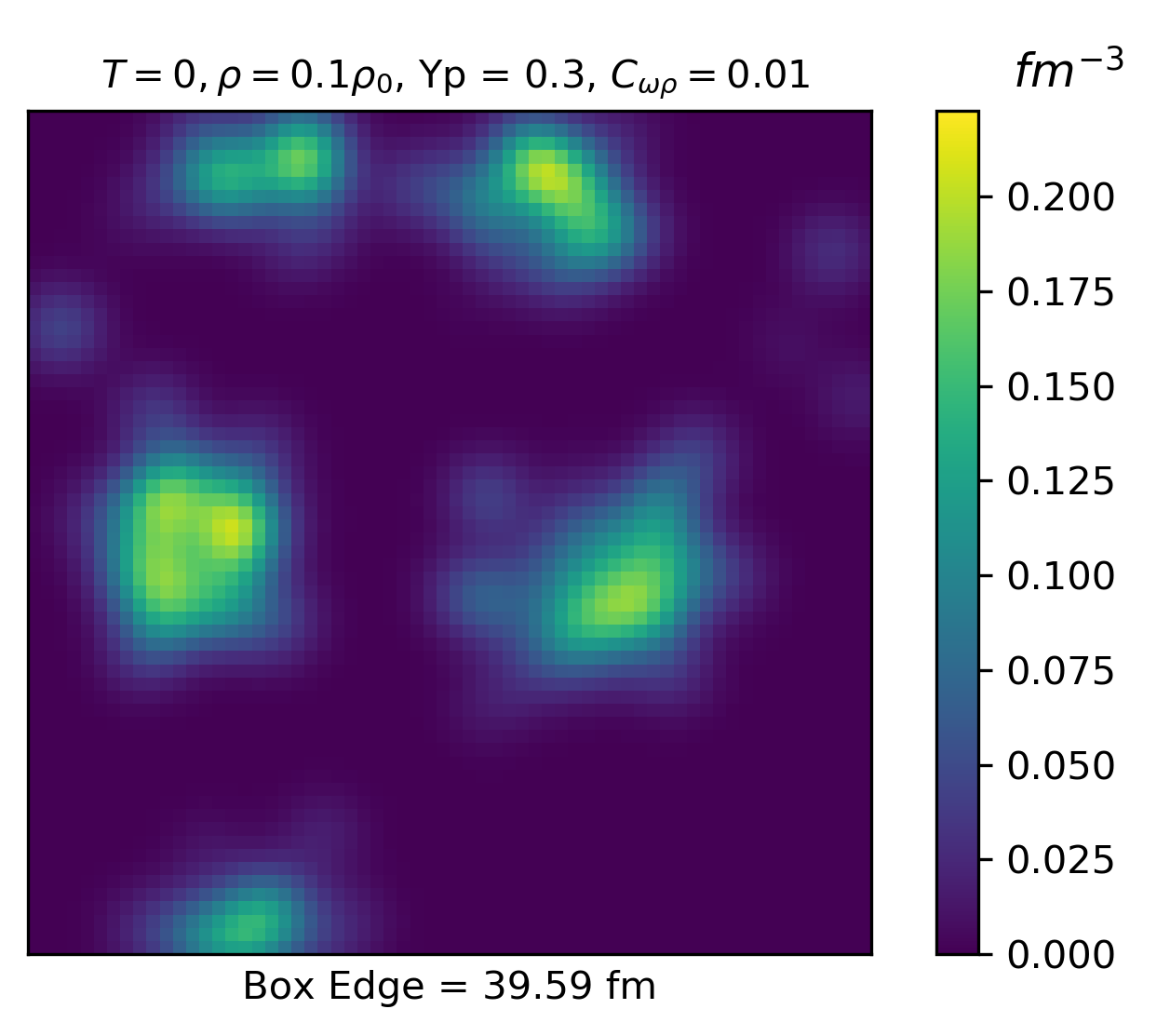}
         \caption{Density map of simulation box with $C_{\omega\rho}$ = 0.01.}
         \label{fig:cluster1}
     \end{figure}
  \vspace{-6pt}  
    \begin{figure}[H]
    \includegraphics[width=7 cm]{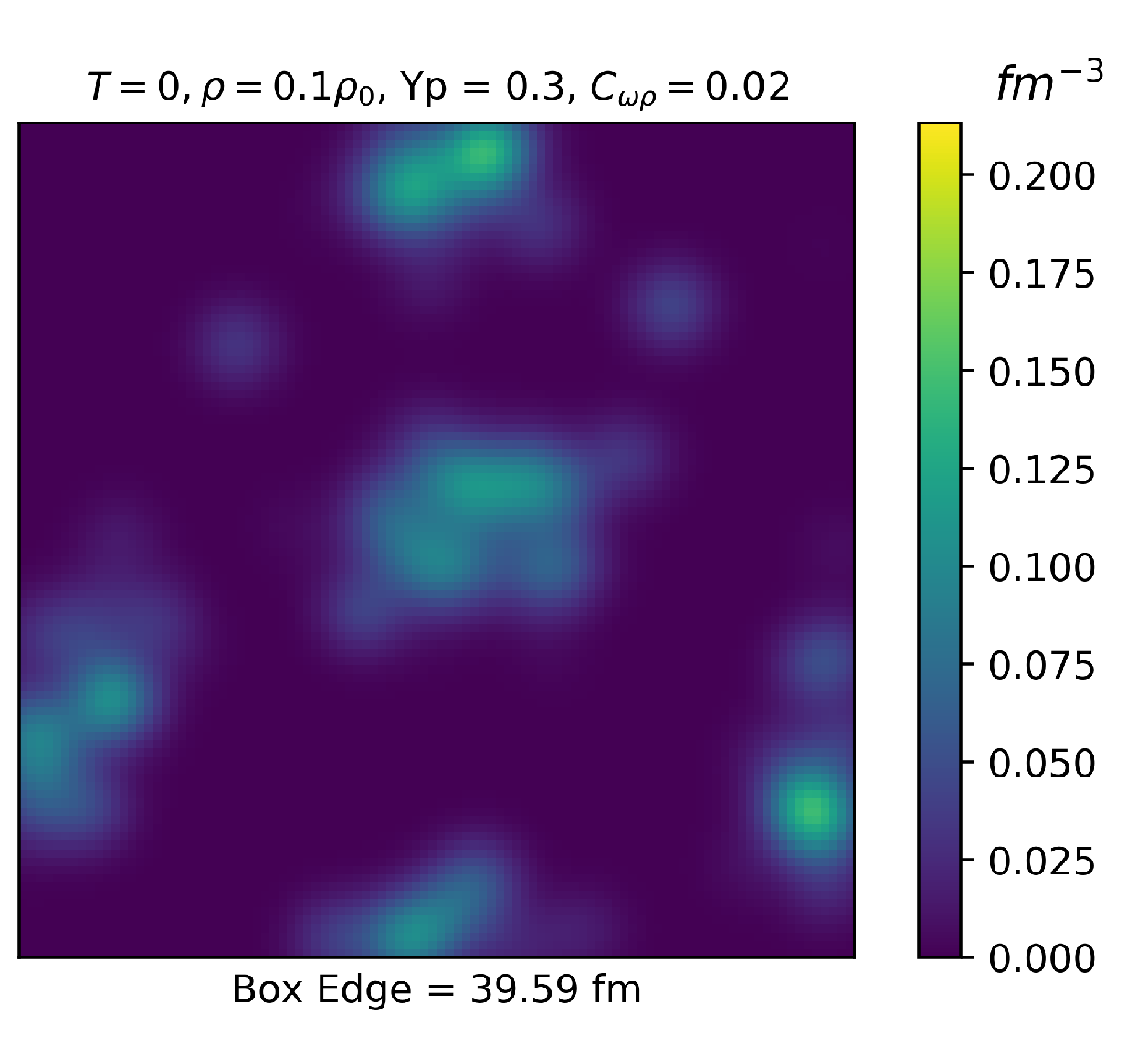}
    \caption{Density map of simulation box with $C_{\omega\rho}$ = 0.02.}
    \label{fig:cluster2}
    \end{figure}
    \vspace{-6pt}
    \begin{figure}[H]
    \includegraphics[width=6.5 cm]{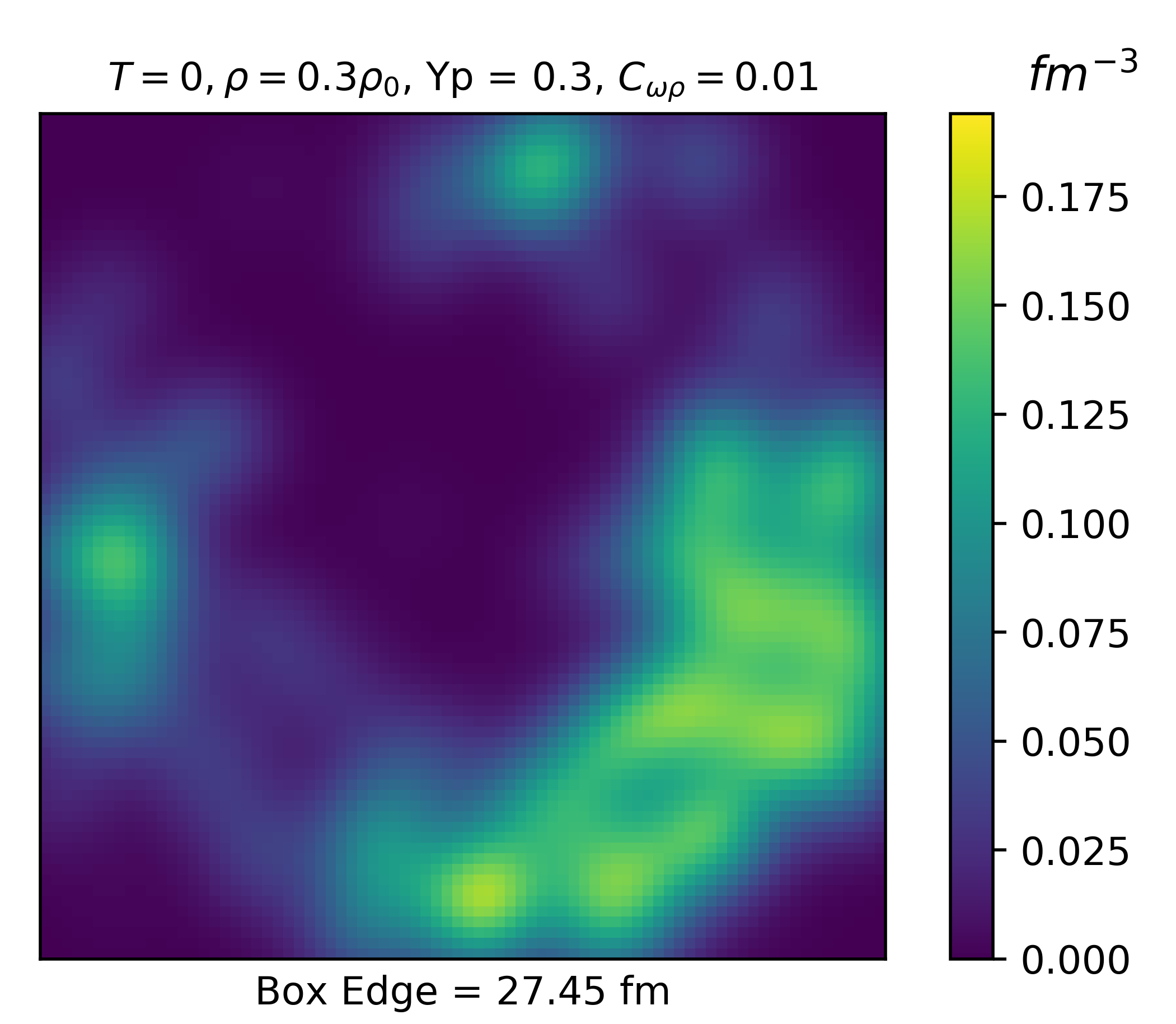}
    \caption{Density map of simulation box with $C_{\omega\rho}$ = 0.01, but~at three times the density of Figure~\ref{fig:cluster1}.}
    \label{fig:cluster3}
    \end{figure}

\subsection{Transition from Clustered to Uniform Nuclear~Matter}

Long-range correlations between nucleons can determine the density at which a liquid-gas phase transition occurs. To~this end, a~useful tool to analyse the spatial distribution of nucleons is the two-point density fluctuation correlation function $\xi_{NN}$ for nucleon density fluctuations defined as~\cite{Nandi2016LowEnergy}:
\begin{equation}\label{eq:twopointcorr1}
\xi_{N N}=\left\langle\Delta_{N}(\mathbf{x}) \Delta_{N}(\mathbf{x}+\mathbf{r})\right\rangle~.
\end{equation}

Here, the~average denoted by $\langle...\rangle$ is taken over the position  $\mathrm{x}$ and in the direction of $\mathrm{r}$. The~fluctuation $\Delta_{N}(\mathrm{x})$ of the nucleon density field $\rho_{N}(\mathrm{x})$ is defined as
\begin{equation}\label{eq:twopointcorr2}
\Delta_{N}=\frac{\rho_{N}(\mathbf{x})-\rho_{\mathrm{av}}}{\rho_{\mathrm{av}}}~,
\end{equation}
where $\rho_{av}$ is the average density of the simulation box. Two-point correlation functions for $Y_P=0.3$ and $0.5$ (for $C_{\omega\rho}$ = 0.01, 0.005, and~$-$0.01) are plotted in Figure~\ref{fig:correlations}. In~all cases, an~increase in density decreases the amplitude of $\xi_{NN}$, indicating a smoother nucleon density distribution. Correlations are highest near the origin as the nucleons have the strongest influence on their nearest neighbors. This also indicates clusterization at low densities. A~negative value of $\xi_{NN}$ at a given $r$ implies anti-clustering or regularity, which means the point at that $r$ has a density lower than the average density of the simulation~box. 


All  curves  at densities higher than $0.8\rho_0$ are almost  flat-lined  at $\xi_{NN} = 0$,  indicating  uniform  matter  above $0.8\rho_0$.  Clear trends in the variation of cluster size and densities with $C_{\omega\rho}$ and $L$ could not be~deduced.

 When nuclear matter is uniform, the~two-point correlation vanishes. At~a certain average density, the~long-range correlations suddenly disappear (instead of gradually), indicating the density turning to uniform matter through a first-order phase transition, which corresponds to the liquid-gas transition. Similar conclusions were obtained in previous studies~\cite{Watanabe2003StructureDynamics,Nandi2016LowEnergy}. For~all cases of $C_{\omega\rho}$ in Figure~\ref{fig:correlations}, more data is needed to find out the point of transition although the transition from asymmetric to uniform matter seems to occur between $\rho/\rho_0$=0.6 and~0.8.  
 
\begin{figure}[H]
    \includegraphics[width=12.5 cm]{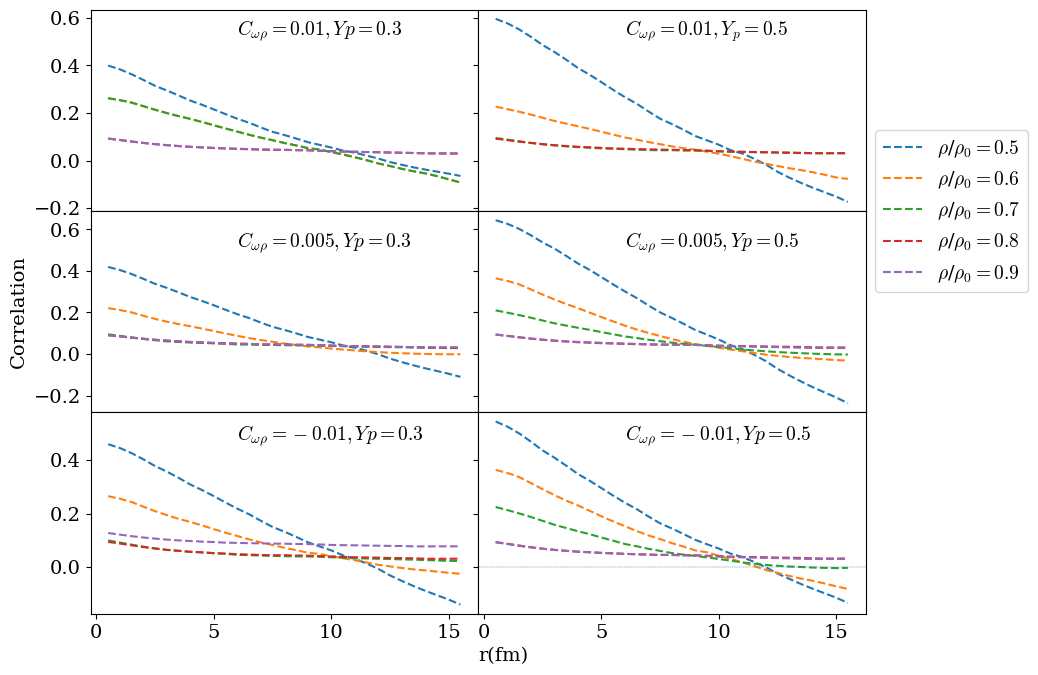}
    \caption{Two-point correlation function $\xi_{NN}$ of nucleon density fluctuations for $C_{\omega\rho}$ = 0.01, \linebreak $C_{\omega\rho}$ = 0.005 and $C_{\omega\rho}$ = $-$0.01 and proton fractions $Y_P$ = 0.3 (\textbf{left}) and 0.5 (\textbf{right}).}
    \label{fig:correlations}
\end{figure}



Note that it has been shown using relativistic mean field models that effects on the slope of the symmetry energy induced by an additional $\omega-\rho$ interaction affect the crust-core transition: a smaller slope reproduces a larger onset~\cite{PhysRevC.103.055812,Ducoin:2011fy,Pais:2016xiu}. Similar results were found within the Brueckner–Hartree–Fock approach~\cite{Vidana:2009is} and in a detailed study involving different approaches~\cite{Pais:2016nzh}.

\section{Conclusions}\label{sec4}
The conditions in the inner crust of neutron stars have been simulated within a Quantum Molecular Dynamics (QMD) approach with periodic boundary conditions to imitate infinite uniform nuclear matter. The~nucleon-nucleon interaction Hamiltonian for QMD was successively developed in earlier works by Aichelin and Stöcker~\cite{Aichelin1986QuantumCollisions}, Peilert~et~al.~\cite{Peilert1991ClusteringDensities}, and~Watanabe~et~al.~\cite{Watanabe2003StructureDynamics} and consists of effective interaction potentials that take into account the Pauli principle, the~Yukawa interaction, Coulomb interaction, and~density dependent terms. In~the current project, an~isospin-dependent potential term to take into account the repulsion from interaction of omega and rho mesons has been implemented in the QMD Hamiltonian. The~idea to include a mesonic self-interaction in nuclear matter calculations is not new. It was first introduced in an attempt to reduce the neutron-skin thickness of \mbox{\ce{^208Pb} \cite{Horowitz2001Neutron208Pb}}. In~a work proposing the IUFSU effective interaction~\cite{Fattoyev2010RelativisticStars}, it was shown that increasing the $\omega-\rho$ coupling constant softens the EoS of nuclear matter at around saturation density and that the density dependence of symmetry energy is highly sensitive to it. This was done within a model based on a relativistic effective field theory. Later, it was shown to improve the radius and tidal deformability of neutron stars, leading RMF models to be in better agreement with observations~\cite{Dexheimer2018WhatGW170817}. 

The new $\omega-\rho$-inspired term in the QMD model Hamiltonian in this work is inspired from the density dependent repulsive Skyrme force and depends on the baryon density and isospin density of asymmetric nuclear matter. A~few values for the coefficient of the $\omega-\rho$ potential were tested, which resulted in very different behavior of the symmetry energy and its slope L. First, the~values and trends of binding energies per nucleon of ground states of several nuclear isotopes were reasonably reproduced compared with experimental values. Simulations for pure neutron matter resulted in a density dependent behavior that is largely similar for all $C_{\omega\rho}$ (coefficient of $\omega-\rho$ meson field interaction) below nuclear saturation density and is in good qualitative agreement within constraints from Chiral EFT. Around $\rho_0$, we can see a large divergence in the trend of $E/N$ and a decrease in maximum energies as $C_{\omega\rho}$ is lowered. The~numerical data obtained by simulating asymmetric nuclear matter was fitted to the energy per nucleon expanded as Taylor series, keeping both the lowest and the second highest order term. The~approximation in the second-order term, also called the parabolic approximation, gave a trend of symmetry energies that decreases as the coefficient of the $\omega-\rho$ potential also decreases. The~corresponding slopes $L$ exhibit a similar trend, although~four of five tested parameter sets produced $L$ values within established constraints~\cite{Li:2019xxz,PhysRevLett.126.172503,PhysRevLett.127.232501}. The~higher-order approximation, which is necessary to obtain a better fitting of the data to energy per nucleon, further reduced the symmetry energies for the same coefficient values whereas there are more variations in the corresponding $L$ values amidst a general decrease. This behavior requires the inclusion of data for higher and in-between values of proton fractions to further improve fitting and obtain better symmetry energy and slope~values. 

The dependence of clusterization in the system, due to the nuclear liquid-gas transition, on~the isospin properties was also explored by calculating two-point correlation functions. Although~a detailed study of the structure of inhomogenous phases could not  be accomplished due to inaccurate Coulomb energies, a~visualization of the simulated system shows interesting pasta-like shapes. The~transition from inhomogeneous to uniform matter is evaluated using a two-point density fluctuation correlation function and  points to a first-order phase transition. Only a small change was observed in the effect of varying $L$ on the transition density, which cannot be deemed as significant. The~properties of the mixed phase with the newly integrated $\omega-\rho$-inspired interaction can be studied in a similar fashion to a work conducted earlier in Ref.~\cite{Nandi2017EffectMatter} with the aim of giving a better range for the critical end-point of the liquid-gas phase transition in dense nuclear matter. The~analysis of two-point density fluctuation correlations also reveals the size of clumps of nucleons in the system. If~the evolution of clump sizes is tracked with respect to time, one can deduce where density fluctuations are amplified enough to have matter separate into domains of high and low densities forming a coexisting phase. The~growth of instabilities or fluctuations point to a region of negative compressibility in the phase diagram of nuclear matter, where at a constant temperature an increase of density results in a decrease in pressure~\cite{Borderie2019LiquidGasNuclei}. This region is called the spinodal region. Therefore, further study is needed to shed more light on the nuclear phase diagram. Steinheimer~et~al.~\cite{Steinheimer2019ACollisions,Steinheimer2014Non-equilibriumState} have conducted detailed analyses on experimental signals of the expected phase transition at large baryon densities and identifying spinodal clumping in high energy nuclear collisions. Studies of temperature, pressure, and time evolution of density fluctuations in nuclear matter are outside the scope of this project but is an interesting prospect for the future. 

QMD has an advantage over other types of models in the possibility to track the trajectory of nucleons and study the non-averaged properties of clusters in nuclear matter at inner crust densities, unlike mean-field approaches. Implementing the $\omega-\rho$ interaction in a QMD model is an important step in efforts to constrain the density dependence of symmetry energy and at the same time observe effects of this interaction on the structure of nuclear matter within a dynamical framework. Pressure can be calculated using the simulated data to obtain the full equation of state, and~subsequently a M-R curve for the model used in this work. We also aim to find a way to reconcile the model with the causality of sound speed, which is ensured in Relativistic Mean Field models but can be problematic in microscopic simulations. An~exciting prospect is finite temperature calculations to check for phases of hot nuclear matter at sub-saturation densities, which is relevant for \mbox{proto-neutron stars.  }    
%
%
%
%
%

%
\vspace{6pt} 

\funding{V. Dexheimer acknowledges support from the National Science Foundation under grant PHY-1748621 and PHAROS (COST Action CA16214). Centre for Scientific Computing (CSC) at the J. W. Goethe-University provided computational support for this project. P. Mehta acknowledges support from the Rolf and Edith Sandvoss-Scholarship awarded by the Walter Greiner Gesellschaft, Frankfurt a.M. 
}

\acknowledgments{P. Mehta would like to thank Horst Stoecker for guidance and enriching discussions. This manuscript is dedicated to the memory of late Stefan Schramm. } 
 }



\begin{adjustwidth}{-\extralength}{0cm}
\reftitle{References}




\end{adjustwidth}

\end{document}